\documentclass[a4paper]{jpconf}
\usepackage{graphicx}
\usepackage{amsmath,amsfonts,amssymb}
\usepackage{marvosym}
\usepackage{txfonts}
\usepackage[T1]{fontenc}
\usepackage{xcolor}
\usepackage{mathtools}
\renewcommand{\d}{\mathrm{d}}

\begin{document}
\title{Dirac QNM spectrum from twisted semiclassical gauge theory of gravity}


\author{Marija Dimitrijevi\'{c} \'{C}iri\'{c}$^1$, Nikola Herceg$^2$, Nikola Konjik$^3$, A. Naveena Kumara$^4$ and Andjelo Samsarov$^5$  }

\address{$^{2,4,5}$ Rudjer Bo\v{s}kovi\'c Institute, Bijeni\v cka  c.54, 10000 Zagreb, Croatia}

\address{$^{1,3}$ Faculty of Physics,  University of Belgrade, Studentski trg 12, 11000 Beograd, Serbia}

\ead{ $^1$dmarija@ipb.ac.rs, $^2$nherceg@irb.hr, $^3$konjik@ipb.ac.rs,  $^4$nathith@irb.hr, $^5$asamsarov@irb.hr}

\begin{abstract}
    Twisted Abelian gauge theory coupled to a noncommutative (NC) Dirac field is studied in order to infer
  the quasinormal mode (QNM) spectrum of the fermion matter perturbations in the vicinity of the  Reissner-Nordstr\"om (RN)  black hole.
The action functional of the theory is invariant under the truncated NC local  $U(1)_{\star}$ gauge transformations
that keep the gravitational background intact. 
The latter, being  a classical gravitational  background unaffected by the 
 NC local  gauge transformations, makes the theory  semiclassical. The most prominent feature of the QNM spectrum is the splitting in the total angular momentum projection 
due to the noncommutativity induced $SO(3) \rightarrow U(1)$ symmetry breaking pattern.
\end{abstract}

\section{Introduction}

The study of Dirac perturbations in the vicinity of a black hole is important for understanding the stability of fermionic matter  in strong gravity regimes. It may also provide some insights  into the behaviour of the Dirac quantum fields in the same regime.
However, the regime of strong gravity is the one where the effects of gravity become comparable 
to the quantum effects,  so much so that the very fabric of spacetime as viewed from the perspective  of classical general relativity comes under serious question. Therefrom arose different approaches  to account for spacetime dynamics in a proper way. These approaches are also closely related to the problem of quantizing gravity, some of them being more fundamental, while others correspond to building effective physical models focused on seizing and describing some of  the  most prominent characteristics of quantum gravity. When it comes to the latter group, noncommutative gauge and gravity theories \cite{Seiberg:1999vs}
 deserve special consideration.
With that in mind, NC gauge and gravity theory, which both fall into a broader framework of NC geometry, can be utilised to construct effective  models of quantum gravity.
In this construction the whole set of physical degrees of freedom or just a part of them  may be considered to be dynamical/noncommutative.
Depending on this, the model constructed within the NC geometry framework is labeled as fully noncommutative or semiclassical. 
Either case (of course, with differing degrees of faithfulness) may be used to infer the properties of the Dirac QNM spectrum
corresponding to perturbations of decaying fermionic matter in the regime of strong gravity or in the presence of  deformed structure of spacetime.

For that purpose, in this brief report we study an effective model of quantum gravity which arises from the
noncommutative gauge theory coupled to NC Dirac field and examine the ensuing fermion perturbations. As both the NC gauge field and the  NC spinor field 
are coupled to the classical gravity background of the Reissner-Nordstr\"om (RN) type, the resulting effective model of quantum gravity is essentially semiclassical.

\section{Semiclassical model of twisted Abelian gauge theory coupled to NC Dirac field} 

We start by introducing  an action functional describing the NC $U(1)_\star$ gauge theory of a  spin-$1/2$ field with charge $q$ on the fixed gravitational background \cite{DimitrijevicCiric:2022ohs}
\begin{equation}    \label{action}
S_\star = \int \d ^4x ~ |e| \star\bar{\hat{\Psi}} \star\Big( i\gamma^\mu \big( \partial_{\mu} \hat{\Psi} - i\omega_\mu \star \hat{\Psi} -iq\hat{A_\mu}\star \hat{\Psi} \big) -m\hat{\Psi}\Big). 
\end{equation}
Noncommutative fields are labeled with a $\hat{\>}$ and the $\star$-product is given by
$ \psi_1 \star \psi_2 =  \mu \circ {\mathcal{F}}\big(  \psi_1 \otimes \psi_2 \big) $
with $\mu$ being  the usual commutative pointwise multiplication of functions, and ${\mathcal{F}}$ the Drinfeld twist operator 
governing the deformation of the usual $U(1)$ gauge theory.
This construction is in line with the usual construction of NC gauge and gravity theories \cite{Jurco:2001rq,Aschieri:2009ky,Aschieri:2011ng,Aschieri:2012in}.
It can be seen that the action (\ref{action})  is invariant under the following infinitesimal $U(1)_\star$ gauge transformations:
\begin{eqnarray}
\delta_\star \hat{\Psi} &=& i\hat{\Lambda} \star \hat{\Psi}, \nonumber  \\
\delta_\star \hat{A}_\mu &=& \partial_\mu\hat{\Lambda} + i  \big( \hat{\Lambda} \star \hat{A}_\mu - \hat{A}_\mu \star \hat{\Lambda}  \big),
  \\
\delta_\star \omega_\mu &=& \delta_\star e^a_{~~\mu} =0, \nonumber
\end{eqnarray}
where $\hat{\Lambda}$ is the NC gauge parameter. As these NC gauge transformations do not affect the gravitational part,
the theory encompassed by the action (\ref{action})  is not completely noncommutative. In this sense, it is semiclassical, characterized by nondynamical gravitational degrees of freedom and fixed gravitational background.
We choose a particular twist of the form
\begin{equation*}
\mathcal{F}=e^{-\frac{i}{2}\theta ^{AB}X_{A}\otimes X_{B}}.  \label{AngTwist}
\end{equation*}
Here $\theta ^{AB}, ~A,B \in \{1,2\}$, are the elements of a constant antisymmetric matrix $\theta ^{12} = - \theta ^{21} = a$ that involve the noncommutative deformation parameter $a$ \cite{Ciric:2017rnf,DimitrijevicCiric:2018blz}.
Moreover, the twist is Abelian, meaning that  $X_{1}=\partial _0$,
$X_{2}= x^1\partial _2 - x^2\partial_1$ are commuting vector fields. This twist satisfies  the conditions of cocyclicity and counitality.
We call it "angular twist" because the vector field $X_{2}= x^1\partial _2 - x^2\partial_1$
is nothing  but a generator of rotations around the $x^3$ direction, that is $X_2 \equiv M_{12} = J_3
=\partial_\phi$. 

The angular twist gives rise to the NC algebra of functions over  ${\mathbb{R}}^4$. In particular,
$ {[ x^0, x^1 ]}_{\star} = -ia x^2, ~~  {[ x^0, x^2 ]}_{\star} = iax^1,$
while all other coordinates commute. These commutation relations are linear in the coordinates; thus, they are of Lie algebra type. 


Gravitational background in the action (\ref{action}) may be specified by any metric that has 
$\partial_\phi$ and $ \partial_t$ as Killing vector fields. Then  the  covariant derivative 
$~ D_{\mu}\hat{\Psi} = {\partial}_{\mu} \hat{\Psi} -i{\omega}_{\mu} \star  \hat{\Psi} - i{\hat{A}}_{\mu}  \star  \hat{\Psi} ~$ transforms as
\begin{equation}
\delta_\star D_\mu \hat{\Psi} = i\hat{\Lambda} \star D_\mu \hat{\Psi}. \nonumber
\end{equation}
The angular twist  in this  case will not act on the gravitational field  and  we will have  $\omega_\mu\star\Lambda = \omega_\mu \cdot \Lambda = \Lambda \star \omega_\mu$.

For simplicity, from now on, we redefine $A_\mu =
qA_\mu$. Then we use the  Seiberg-Witten (SW) map \cite{Seiberg:1999vs,Jurco:2001rq} to express  the NC fields $ \hat{\Psi}$ and $\hat{A}_\mu$ as functions of the corresponding commutative fields
and deformation parameter $a$. The SW map assumes an expansion  in the deformation parameter  and this expansion is known to all orders for an arbitrary Abelian twist deformation, of which the angular twist is only one example.
For  the angular twist operator that we consider, SW map gives rise to the following expansions for the fields:
\begin{eqnarray}
\hat{\Psi} &=& \Psi -\frac{1}{2}\theta^{\rho\sigma}A_\rho(\partial_\sigma\Psi) , \label{HatPsi}   \nonumber  \\
\hat{A}_\mu &=& A_\mu -\frac{1}{2}\theta^{\rho\sigma}A_\rho(\partial_\sigma A_{\mu} +
F_{\sigma\mu}).  \nonumber
\end{eqnarray}
The expanded action  up to the first order in the deformation parameter $a$  is given by
\begin{eqnarray}
  S_\star &=& \int \d ^4x ~ |e|  \Bigg[  \bar{\Psi} \Big( i\gamma^\mu D_\mu \Psi -m\Psi\Big) \nonumber \\
&& +\frac{1}{2}\theta^{\alpha\beta}\Big( -iF_{\mu\alpha}\bar{\Psi}\gamma^\mu D^{\mbox{\tiny{U(1)}}}_\beta\Psi
-\frac{i}{2}\bar{\Psi}\gamma^\mu \omega_\mu F_{\alpha\beta}\Psi 
  -\frac{1}{2}F_{\alpha\beta}\bar{\Psi} \big(i\gamma^\mu D^{\mbox{\tiny{U(1)}}}_\mu \Psi -m\Psi\big) \Big) \Bigg].
\end{eqnarray}
Finally,  the gravitational background is fixed to be that of a charged non-rotating black
hole in $4$ dim, the Reissner-Nordstr$\ddot{o}$m black hole. The RN metric tensor in spherical coordinates is given by
\begin{equation}
g_{\mu\nu}=
\frac{\Delta}{r^2} {\d t}^2  -\frac{r^2}{\Delta} {\d r}^2 - r^2 {\d \theta}^2  -  r^2  {\sin^2 \theta}{\d \phi}^2,\nonumber
\end{equation}
where $\Delta = r^2 - 2MGr+ Q^2 G$ and $M$ and $Q$ are, respectively, the mass and the charge of the RN black hole.

The vierbein frame  can be chosen as
\begin{equation} \label{metrictetrad}
  e^a_{~\mu} =
\left( \begin{array}{ccccc}
  \frac{\sqrt{\Delta}}{r} & 0  & 0 & 0  \\
   0   & \frac{r}{\sqrt{\Delta}} & 0 & 0  \\ 
   0  & 0 & r &  0 \\
 0  & 0 & 0 &  r \sin \theta \\
\end{array} \right),  \quad \quad 
 e_a^{~\mu} =
\left( \begin{array}{ccccc}
  \frac{r}{\sqrt{\Delta}} & 0  & 0 & 0 \\
   0   & \frac{\sqrt{\Delta}}{r} & 0 &  0  \\ 
   0  & 0 &  \frac{1}{r} & 0 \\
  0  & 0 & 0 &  \frac{1}{r \sin \theta }  \\
\end{array} \right)    \nonumber
\end{equation}
with  the following representation of gamma matrices
\begin{equation} 
  \gamma^0 = i  \tilde{\gamma}^0 =
 i \left( \begin{array}{ccccc}
  0  & I  \\
   I   & 0  \\ 
\end{array} \right),  \quad \quad 
 \gamma^1 =   i \tilde{\gamma}^3 =
i \left( \begin{array}{ccccc}
  0  & \sigma_3  \\
   -\sigma_3   & 0  \\  
 \end{array} \right),   \nonumber
\end{equation} 
\begin{equation} 
 \gamma^2 =   i \tilde{\gamma}^1 =
i\left( \begin{array}{ccccc}
  0  & \sigma_1  \\
   -\sigma_1   & 0  \\ 
\end{array} \right),  \quad \quad 
 \gamma^3 =   i \tilde{\gamma}^2 =
i\left( \begin{array}{ccccc}
  0  & \sigma_2  \\
   -\sigma_2   & 0  \\ 
\end{array} \right),     \nonumber
\end{equation} 
where $\tilde{\gamma}^0$,  $\tilde{\gamma}^1$, $\tilde{\gamma}^2$ and $\tilde{\gamma}^3$ are gamma matrices in chiral/Weyl representation, while  $~ \sigma_i, ~ (i=1,2,3)~$ are the usual Pauli matrices.

On the other hand, the gauge field part is also fixed by the RN background. The latter, being non-rotating, gives rise to the gauge field $A_\mu$ and the field strength $F_{\alpha\beta}$, whose non-zero components are
\begin{equation*}
A_t = -\frac{qQ}{r}, \qquad 
  F_{rt} = \frac{qQ}{r^2}.
\end{equation*}
This leads to a simplified NC action
\begin{equation}
  S_\star = \int \d ^4x ~ |e| \Bigg[ \bar{\Psi} \Big( i\gamma^\mu D_\mu \Psi -m\Psi\Big) -\frac{i}{2}\theta^{\alpha\beta}\bar{\Psi} F_{\mu\alpha}\gamma^\mu D^{\mbox{\tiny{U(1)}}}_\beta\Psi \Bigg]. 
\end{equation}
Mathematically, the semiclassical approximation here manifests itself  in the following way: the covariant derivative $D_\mu \Psi = \partial_\mu\Psi - iA_\mu \Psi - i\omega_\mu \Psi$ includes both the electromagnetic  $(U(1)) $  and the gravitational part, while the covariant derivative $D^{\mbox{\tiny{U(1)}}}_\beta\Psi = \partial_\beta\Psi - iA_\beta \Psi$ has only the electromagnetic part. In the NC correction, only the $U(1)$  part appears.

In addition, the only non-zero components of $\theta^{\alpha\beta}$ are $\theta^{t\phi}=
-\theta^{\phi t}=a$. Putting these remarks together and  including the explicit expression for  $\gamma^{r} = e_a^{~~r} \gamma^{a},$ 
 the equation of motion for the spinor field  $\Psi$ reduces to
\begin{equation}     \label{EoMPsiNC} 
   i\gamma^\mu \Big(\partial_\mu \Psi -i \omega_\mu\Psi - iA_\mu\Psi\Big) -m\Psi -
    \frac{ia}{2}  \frac{qQ}{r^2}  \frac{\sqrt{\Delta}}{r} \gamma^1 \partial_{\phi} \Psi = 0 . \nonumber
\end{equation}
 Inserting the vierbein frame  with the gamma matrices in the Weyl representation  and    writing the  equation in terms of  the  two-component spinors
 $\Psi =  \left(\begin{matrix} \Psi_1  \\ \Psi_2  \\   \end{matrix}\right)$ yields \cite{Herceg:2025zkk}
\begin{equation*} \label{easiercomparison}
\begin{split}
i\frac{r}{\sqrt{\Delta}} i 
\begin{pmatrix} 0 & \mathbf{1} \\ \mathbf{1} & 0 \end{pmatrix} \partial_t \Psi
+ i \frac{\sqrt{\Delta}}{r} i 
\begin{pmatrix} 0 & \sigma_3 \\ -\sigma_3 & 0 \end{pmatrix} \partial_r \Psi
+ i\frac{1}{r} i 
\begin{pmatrix} 0 & \sigma_1 \\ -\sigma_1 & 0 \end{pmatrix} \partial_\theta \Psi
+ i\frac{1}{r\sin\theta} i 
\begin{pmatrix} 0 & \sigma_2 \\ -\sigma_2 & 0 \end{pmatrix} \partial_\phi \Psi \\
+ \Big( e_0^{~~t} \gamma^0 \omega_t + e_2^{~~\theta} \gamma^2 \omega_\theta
+ e_3^{~~\phi} \gamma^3 \omega_\phi \Big)\Psi
+ e_0^{~~t} \gamma^0 A_t \Psi
- m\Psi
+ \frac{a}{2}\frac{qQ}{r^2} \frac{\sqrt{\Delta}}{r}
\begin{pmatrix} 0 & \sigma_3 \\ -\sigma_3 & 0 \end{pmatrix} \partial_\phi \Psi
= 0.
\end{split}
\end{equation*}
The separation of the equation is achieved with the ansatz \cite{Herceg:2025zkk}
\begin{equation}  
\Psi =
  e^{i(\nu \phi - \omega t)}   \left(\begin{matrix} \psi_1 (r, \theta)  \\ \psi_2 (r, \theta )  \\ \end{matrix}\right)  =
   e^{i(\nu \phi - \omega t)}  \left(\begin{matrix}     - r^{-1/2} \Delta^{1/4} \xi_{+\frac{1}{2}} (r) S_1 (\theta) \\- r^{-1/2} \Delta^{-1/4} \xi_{-\frac{1}{2}} (r) S_2 (\theta)  \\  
     r^{-1/2} \Delta^{-1/4} \xi_{-\frac{1}{2}} (r) S_1 (\theta)  \\ r^{-1/2} \Delta^{1/4} \xi_{+\frac{1}{2}} (r) S_2 (\theta)  \\ \end{matrix}\right),   \nonumber
\end{equation}
with $\xi_s, s \in \{ +\frac{1}{2}, -\frac{1}{2} \},$  describing the radial part  that satisfies
\begin{eqnarray}   \nonumber
 \Delta \partial_r^2 \xi_s &+& \Bigg(  2(s + 1) (r - M)  - i \nu a qQ f -  2s \frac{m\Delta}{\lambda_s + 2s mr}  \Bigg) \partial_r \xi_s  \nonumber \\
&+&\Bigg[ \frac{{( \omega r^2 - qQr )}^2  - 2 i s(r - M)( \omega r^2 - qQr ) }{\Delta} + 4i s \omega r - 2 i s qQ  - \lambda_s^2  \Bigg] \xi_s  \nonumber  \\
&-&\Bigg[ \frac{i\nu a qQ }{r^3}  \Big(  s r^2  +  (1- s) Mr  - Q^2 \Big) + \frac{m}{\lambda_s + 2s mr} \Bigg(2s (s + \frac{1}{2}) (r - M)   \nonumber  \\
&+&  i \omega r^2 - iqQr - 2s\frac{i\nu aqQ}{2} f   \Bigg)  - m^2 r^2  \Bigg] \xi_s  =0.    \nonumber
\end{eqnarray}
In the above, $\lambda_s$ is a separation constant satisfying $ \lambda_s^2 = (j - s)(j + s + 1).$  
In the following, we will consider massless perturbations $(m=0).$  Introducing the tortoise coordinate $y$ satisfying $d y / d r = r^2 \Delta^{-1} \big( 1+ ia\nu qQ/r \big)^{-1},$
i.e.
\begin{eqnarray} \label{modtortoise1}
y &=&  r_*^{RN} -ia \nu qQ  ~ \Bigg \{  \frac{r_+}{r_+ - r_-} \ln (r- r_+) - \frac{r_-}{r_+ - r_-} \ln (r- r_-) \Bigg \}  ,\nonumber
\end{eqnarray}
with  $ r_*^{RN} $ being the standard tortoise coordinate for the Reissner-Nordstr\" {o}m metric, 
and making the field  transformation  $\chi_s (r) = \Delta^{s/2} r \xi_s (r) $,  leads to the fermion perturbation equation in
the Schr\"odinger form
\begin{equation} \label{schrod2}
\frac{\d^2 \chi}{\d {y}^2}  + V \chi =0.     \nonumber
\end{equation}
The effective potential $V$ is given by
\begin{eqnarray} \label{veffektive}
 V &=& \frac{\Delta}{r^4} \Bigg[ \frac{2Q^2}{r^2} - \frac{2M}{r} - j(j+1) + s^2 
    +\frac{{  \big( \omega r^2 - qQr  - i s(r - M) \big) }^2   }{\Delta} + 4i s \omega r   \nonumber  \\
 &-& 2 i s qQ
 + \frac{ i a \nu  qQ \Delta}{r^3}     
+   is a \nu qQ \frac{r - M}{r^2}  - \frac{ i a \nu  qQ }{r^3}  \bigg( s r^2 + (1-s) Mr - Q^2 \bigg)    \nonumber \\
& +& 2i a \nu \frac{qQ}{r} \Big(  \frac{2Q^2}{r^2} - \frac{2M}{r} - j(j+1) + s^2   \Big)     
+ 2i a \nu \frac{qQ}{r} \frac{{ \big( \omega r^2 - qQr  - i s(r - M) \big) }^2   }{\Delta}  \nonumber \\
 &-&    8s a \nu  \omega qQ + 4 s a \nu  \frac{q^2 Q^2}{r} \Bigg].    
\end{eqnarray}

\section{Continued fraction method and results for the Dirac QNM spectrum}

Next we implement the continued fraction method \cite{leaver,nollert} in order
to determine the QNM spectrum for a massless charged fermion field around the RN black
hole in the presence of  noncommutative deformation of spacetime. This method is one of the more robust and less restrictive ones.  In many cases it is applicable to a wide range of system parameters.

The asymptotic form of the quasinormal modes which takes into account QNM boundary conditions is given by
\begin{gather}    
  \xi_s (r) \rightarrow 
\begin{cases}         
  Z_{\text{out}} e^{i \omega y} y^{-1-i  qQ  - 2s -  a\nu qQ \omega}, & \text{for } r \rightarrow \infty,\>\> (y\rightarrow \infty) \\  
  \\
  Z_{\text{in}} \frac{1}{{(r - r_+)}^{s/2}}  e^{-i \Big( \omega  - \frac{ qQ}{r_+} - is \frac{r_+ - r_-}{2r_+^2} \Big)  \Big( 1 + ia \nu  \frac{ qQ}{r_+}  \Big)y },   & \text{for }r \rightarrow  r_+, \>\> (y\rightarrow -\infty)                
\end{cases} , \nonumber
\end{gather}
where 
$$
  y = r + \frac{r_+}{r_+ - r_-} \Big(r_+ - iam qQ \Big) \ln (r- r_+) - \frac{r_-}{r_+ - r_-} \Big(r_- - iamqQ \Big) \ln (r- r_-)  
$$  
is the tortoise coordinate for the case in hand and $Z_{\text{out}}$ and  $Z_{\text{in}}$ are the constant amplitudes of the outgoing and ingoing waves, respectively.

The perturbation equation  has an irregular singularity at $r=+\infty$ and three regular singularities at
$r=0$, $r=r_-$ and $r=r_+$. In order to apply Leaver's method, one expands the general solution  in terms of power series
around  $r = r_+$. Then the radial part of the spin $1/2$  field takes the form 
\begin{equation}  \label{generalpowersolution}
 \xi_s (r) = e^{i \omega r}  {(r-r_-)}^{\epsilon} \sum_{n=0}^{\infty} a_n {\Big( \frac{r-r_+}{r-r_-} \Big)}^{n + \delta}. \nonumber
\end{equation}
The parameters $\epsilon$ and $\delta$ are given by
\begin{equation}  \label{epsilondelta}
\delta = -i \frac{r_+^2}{r_+ - r_-} \Big( \omega - \frac{qQ}{r_+} \Big) - s, \qquad
\epsilon =   i\omega (r_+ + r_- ) -1 -2s  - i qQ .   \nonumber
\end{equation}
We first set $a=0$ and $m = 0$. This corresponds to
an undeformed (commutative) (un)charged massless fermion  field in the RN background.  The   analysis of  the corresponding QNM spectrum by the continued fraction method has been carried out in \cite{Richartz:2014jla,chowdhury}.
   The problem is reduced to the following  $3$-term recurrence relations
\begin{eqnarray}    
  \alpha_n a_{n+1} + \beta_n a_n +\gamma_n a_{n-1} = 0, \nonumber \\
   \alpha_0 a_{1} + \beta_0 a_0   = 0,       \nonumber
\end{eqnarray}
where the coefficients $\alpha_n, \beta_n$ and $\gamma_n$ are given as
\begin{align}
  \alpha_n  &=  -(n+1) \Big( r_-(n-s+1)+r_+(-n+s-1-2iqQ+2ir_+\omega) \Big), \label{contfrsimple} \nonumber  \\
    \beta_n  &=   -r_+\Big(\lambda_s+2n^2-4ir_+\omega(2n+1+3iqQ)+6inqQ+2n-4(qQ)^2+3iqQ   \nonumber\\
  & \quad -8r_+^2\omega^2+s+1 \Big)  + r_- \big(\lambda_s+2n(n+1+iqQ)+iqQ+s+1 \big)-2i(2n+1)r_+r_-\omega,  \nonumber   \\
  \gamma_n &= -\Big(n+2i\big(qQ-\omega(r_++r_-)\big)\Big)\Big(n(r_--r_+)+ir_+(-2qQ+2r_+\omega+is)+r_-s\Big).  \nonumber
\end{align}
In a general case when $a \neq 0$ (and $m = 0$), the spacetime deformation gives rise to the $6$-term recurrence relations
\begin{eqnarray}  \label{6contfr}
A_n a_{n+1} + B_n a_n +C_n a_{n-1} + D_n a_{n-2} + E_n a_{n-3} + F_n a_{n-4 }  &=& 0, \quad  n\geqslant 4\nonumber \\
A_3 a_{4} + B_3 a_3 +C_3 a_{2} + D_3 a_{1} + E_3 a_{0}   &=& 0, \quad n=3\nonumber \\
A_2 a_{3} + B_2 a_2 +C_2 a_{1} + D_2 a_{0}   &=& 0, \quad n =2    \nonumber  \\
A_1 a_{2} + B_1 a_1 +C_1 a_{0}    &=& 0, \quad n=1\nonumber \\
A_0 a_{1} + B_0 a_0   &=& 0, \quad n=0.  \nonumber
\end{eqnarray}
where the coefficients $A_n, B_n, C_n, D_n, E_n $ and $ F_n$ are given by
\begin{align}   
  A_n   &=   r_+^3 \alpha_{n},  \nonumber \\
  B_n  &=  r_+^3 \beta_n - 3 r_+^2 r_- \alpha_{n-1}  -ia\nu qQr_+\Big(\frac{r_+-r_-}{2}+(n-s)(r_+-r_-)-ir_+(\omega r_+-qQ)+(r_+-r_-)\frac{s}{2}\Big), \nonumber \\
  C_n  &=  r_+^3 \gamma_n  + 3r_+ r_-^2 \alpha_{n-2}  -3r_+^2 r_- \beta_{n-1} +a\nu qQ\omega r_+(r_+-r_-)^3  
 -ia\nu qQ(r_+-r_-)^2 \big( -1-2s-iqQ+i\omega (r_++r_-) \big) r_+   \nonumber\\
 & + ia\nu qQ(r_+-r_-)(2r_++r_-)\big((n-1-s)(r_+-r_-) - ir_+(\omega r_+-qQ)\big) +          ia\nu qQ  {(r_+ - r_-)}^2  \big( r_+ -  \frac{1}{2} (1 - s) r_- \big),\nonumber \\
 D_n  &= - r_-^3 \alpha_{n-3}  + 3r_+ r_-^2 \beta_{n-2} -3 r_+^2 r_- \gamma_{n-1}  
   +ia\nu qQ(r_+-r_-)^2(r_++r_-)\big(-1-2s-iqQ+i\omega (r_++r_-)\big) \nonumber\\
&   - ia\nu qQ(r_+-r_-)(2r_++r_-)\big((n-2-s)(r_+-r_-)-ir_+(\omega r_+-qQ)\big)
  -ia\nu qQ(r_+ - r_-)^3 (1 - i \omega r_-)    \nonumber\\
&+ \frac{1}{2} i a\nu qQ (1 + s) r_+  {(r_+ - r_-)}^2 ,  \nonumber \\
  E_n  &=   3r_+ r_-^2 \gamma_{n-2} - r_-^3 \beta_{n-3} -ia\nu qQ(r_+-r_-)^2\frac{r_-}{2}-ia\nu qQ(r_+-r_-)^2\big(-1-2s-iqQ+i\omega (r_++r_-)\big)r_- \nonumber\\
&+ia\nu qQ(r_+-r_-)r_-\big((n-3-s)(r_+-r_-)-ir_+(\omega r_+-qQ)\big) + \frac{1}{2} ia\nu qQ s r_+ {(r_+ - r_-)}^2,
 \nonumber\\
  F_n &= -r_-^3 \gamma_{n-3}\nonumber \\ \nonumber
\end{align}
Solving these relations requires  three consecutive applications of the Gaussian elimination method \cite{DimitrijevicCiric:2019hqq} that result in the more familiar $3$-term recurrence relation.
The third and the last Gaussian elimination  leads to the required $3$-term recurrence relation 
\begin{eqnarray*}    \label{contfrsimplegauss}
  A_n^{(3)} a_{n+1} + B_n^{(3)} a_n +  C_n^{(3)} a_{n-1} = 0, \nonumber \\
   A_0^{(3)} a_{1} + B_0^{(3)} a_0   = 0.  
\end{eqnarray*}
where the  coefficients of the third level,  $A_n^{(3)}, B_n^{(3)}, C_n^{(3)}$ (obtained after the final step in the three-stage Gaussian elimination process), are not accessible in an explicit form, but are given
by a special iterative algorithm \cite{DimitrijevicCiric:2019hqq}.
 In contrast to the recurrence relations with a number of terms higher than three, the algorithm  for solving any $3$-term recurrence relation is well known and the fundamental QNM frequencies 
in our case will follow by finding the roots of the following continued fraction
\begin{equation}   \label{infcontfracgauss}
0= B_0^{(3)} - \cfrac{A_0^{(3)} C_1^{(3)}}{B_1^{(3)} - \cfrac{A_1^{(3)} C_2^{(3)}}{B_2^{(3)} -\cfrac{A_2^{(3)} C_3^{(3)}}{ B_3^{(3)} - \cdot \cdot \cdot  \cfrac{A_n^{(3)} C_{n+1}^{(3)}}{B_{n+1}^{(3)} - \cdot \cdot \cdot}  }}} \quad .
\end{equation} 
 However, as the continued fraction represents an infinite series, it must be truncated to a finite number of terms to make the numerical evaluation possible. In our numerical computations, we have taken $N$ as high as $N \sim 200$. 
To increase the convergence rate and enhance accuracy, we apply Nollert's  method  \cite{nollert}.  The Nollert method takes into account that the quantity $R_N=-a_{N+1}/a_N$ complies with the  relation
\begin{equation}
\label{nolrel}
R_N=\frac{C^{(3)}_{N+1}}{B^{(3)}_{N+1} - A^{(3)}_{N+1}R_{N+1}}.
\end{equation}
The quantity $R_N$ accurately represents the contribution from the truncated tail of the continued fraction (\ref{infcontfracgauss})
evaluated at order $N$. If $R_N$   is expanded as 
$ R_N= \sum_{k=0}^{\infty} \Xi_k N^{-k/2}$ and inserted  into equation \eqref{nolrel},  
the coefficients $\Xi_k$ may be obtained in an explicit form. Since for all practical purposes the coefficients $ A^{(3)}_{N+1}$, $ B^{(3)}_{N+1}$, and $ C^{(3)}_{N+1}$ can  be evaluated only numerically using the Gaussian elimination method, we take the expressions for the commutative values of $\Xi_k$ provided in the literature \cite{Richartz:2014jla}.



In the rest of this report we analyze  the effects of noncommutative deformation on the Dirac QNM spectrum in our model and present some of the more notable features. For that purpose let us first recall that the parameter  $a$ controls the NC deformation and $\nu = -j,\ -j +1,..., j$ is the projection of the total angular
momentum $j$. As the projection $\nu$ in the effective potential  (\ref{veffektive})
is always coupled to the deformation $a,$ a splitting of the QNM frequencies in $\nu$ is  expected
to emerge as a genuine effect of noncommutative deformation. That this will indeed be the case is demonstrated
with the figures presented below.


The first pair of figures shows
the dependence of the fundamental QNM frequency $\omega = {\mbox Re}\, \omega +i {\mbox Im}\,
\omega$ on the charge $qQ$ for the fermion  field in the channel  ($j=3/2$, $s=1/2$).   The remaining parameters are fixed as follows:  $Q=0.5$, $a=0.1$, and $M=1$, which amounts to the non-extremal case with the extremality $Q/M = 0.5$.
 The splitting in QNM frequencies is  clearly illustrated for their real (highlighted in the inset) and imaginary parts  for various magnetic quantum numbers $\nu$. 

\begin{center}
\begin{tabular}{lll}
\includegraphics[scale=0.25]{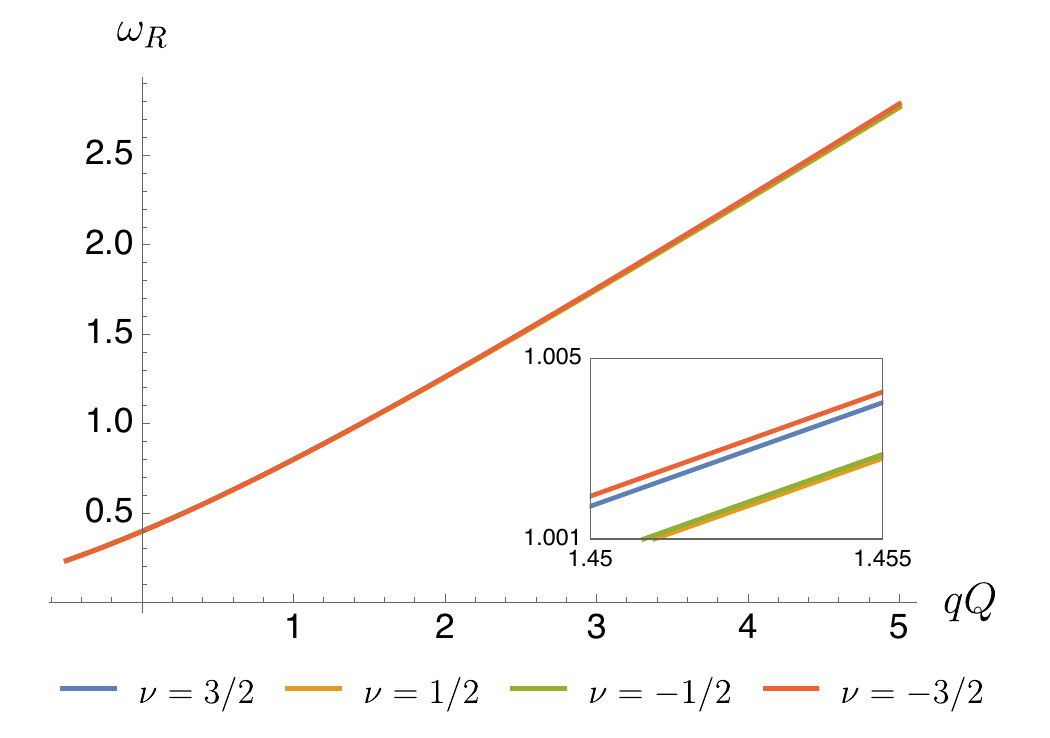}& & \hspace{-0mm}\includegraphics[scale=0.25]{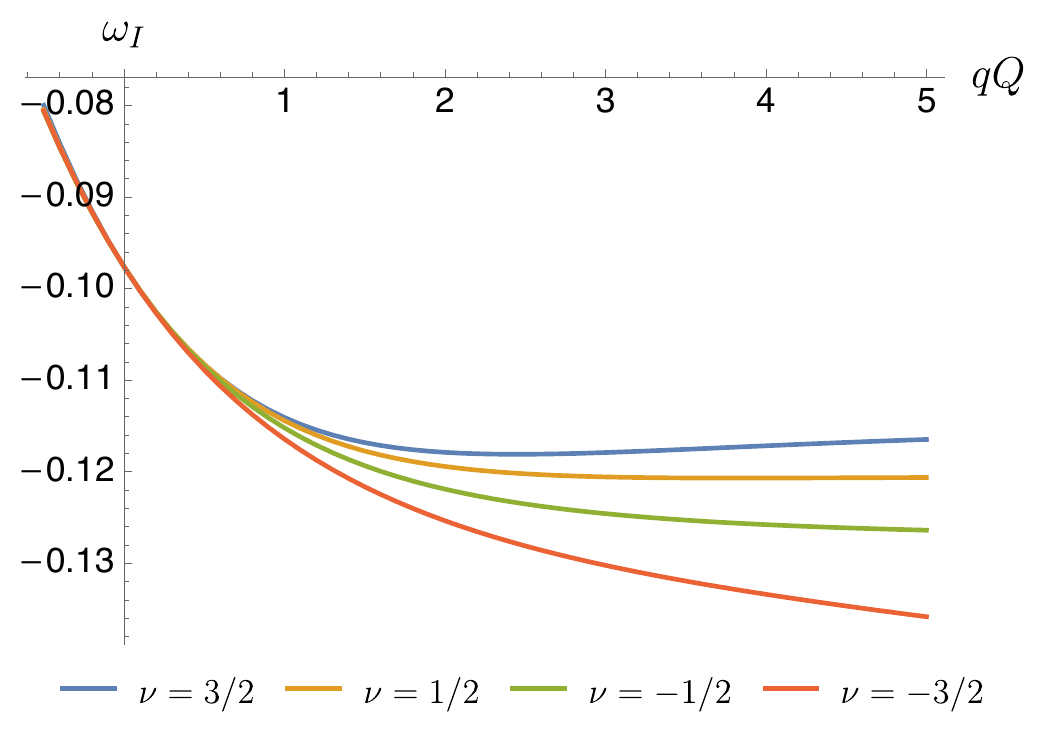}
\end{tabular} 
\end{center}


The second pair of figures  illustrates the mere effect of deformation by directly confronting the noncommutative values for the QNM frequencies with  the commutative ones. It is expressed through the differences between NC and commutative QNM frequencies, $(\omega^{NC}-\omega^{C})$, as functions of $qQ$ for the real and imaginary parts, respectively. The channel 
considered is  ($j=3/2$, $s=1/2$) and the remaining parameters are the same as before, with the extremality $ Q/M =0.5$.
 Note that the splitting here is nonsymmetric in the projection $\nu$,  in contrast to the case with noncommutative scalar field \cite{DimitrijevicCiric:2019hqq}.

\begin{center}
\begin{tabular}{lll}
\includegraphics[scale=0.25]{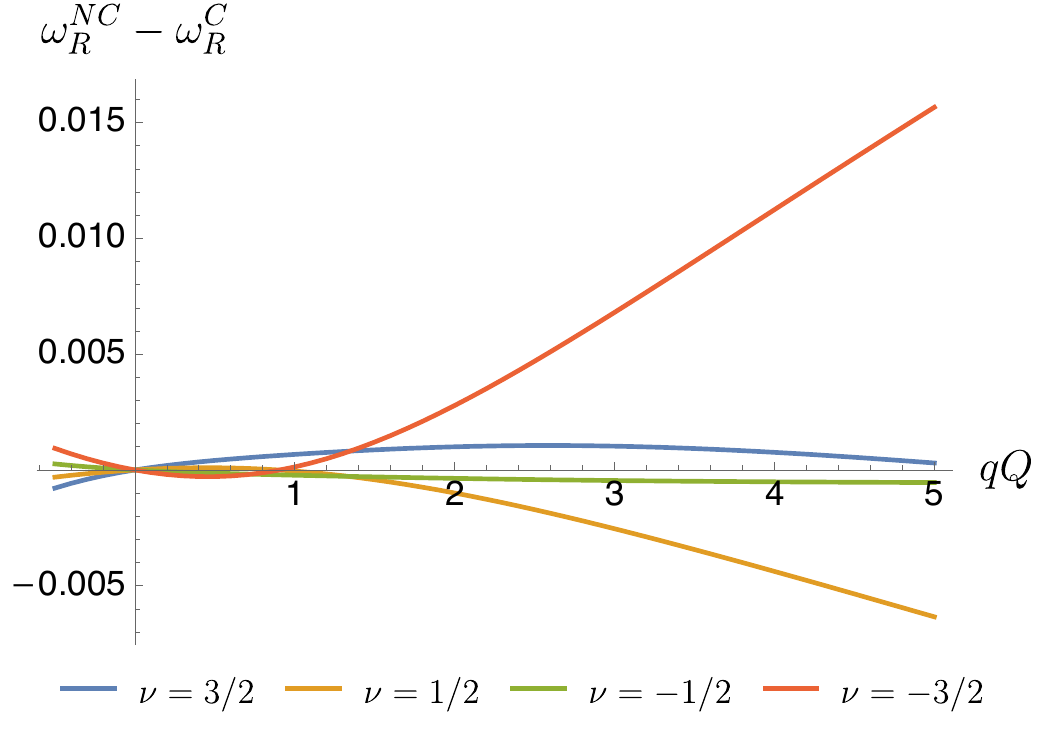}& & \hspace{-0mm}\includegraphics[scale=0.25]{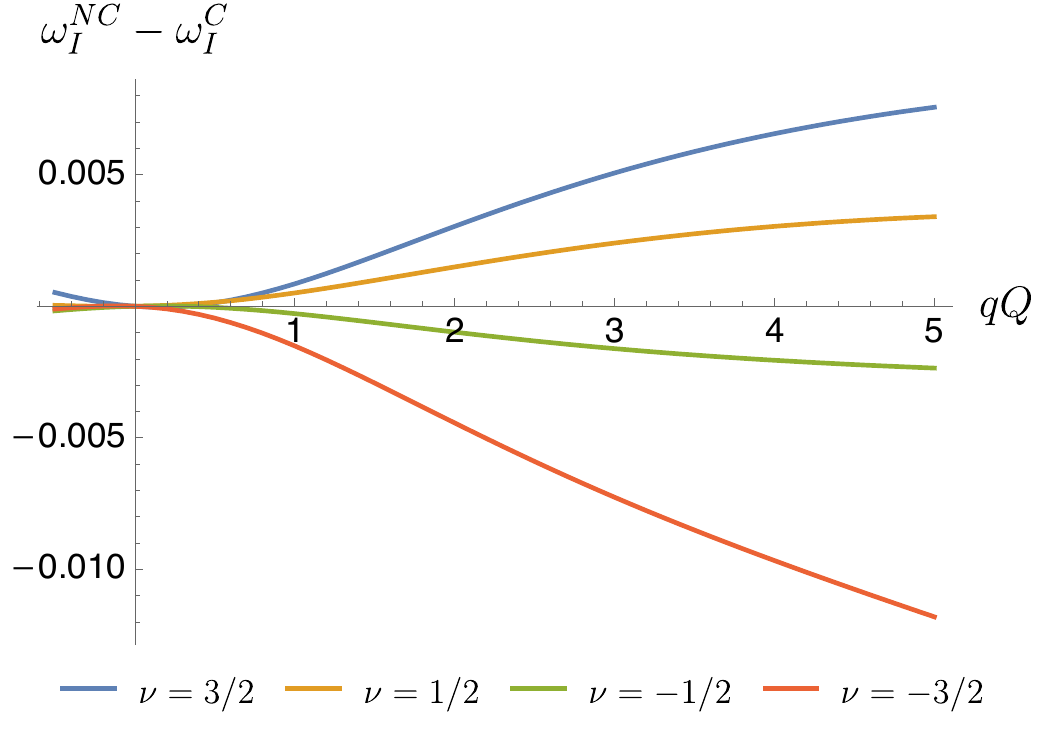}
\end{tabular} 
\end{center}

The third pair of figures illustrates
the dependence of the real and the imaginary part of the NC QNM frequencies on the fermion field charge $q$
for the channel $(j = 3/2, s = 1/2)$ and for all three total angular momentum projections in that channel,
$\nu \in \{-3/2, -1/2, 1/2, 3/2\}$. 
On the same figures, the dependence of $\omega_R$ and $\omega_I$ versus the fermion field charge $q$ is also  shown for different extremalities $Q/M$,
with  different extremalities being  shown in different colors.
The value of $Q/M$ varies from $0.1$ up to near extremal value of $0.99$. Although the splitting in $\nu$ is not visible on the left panel, it is clearly present  on the right one. 
\begin{center}
\begin{tabular}{lll}
\includegraphics[scale=0.25]{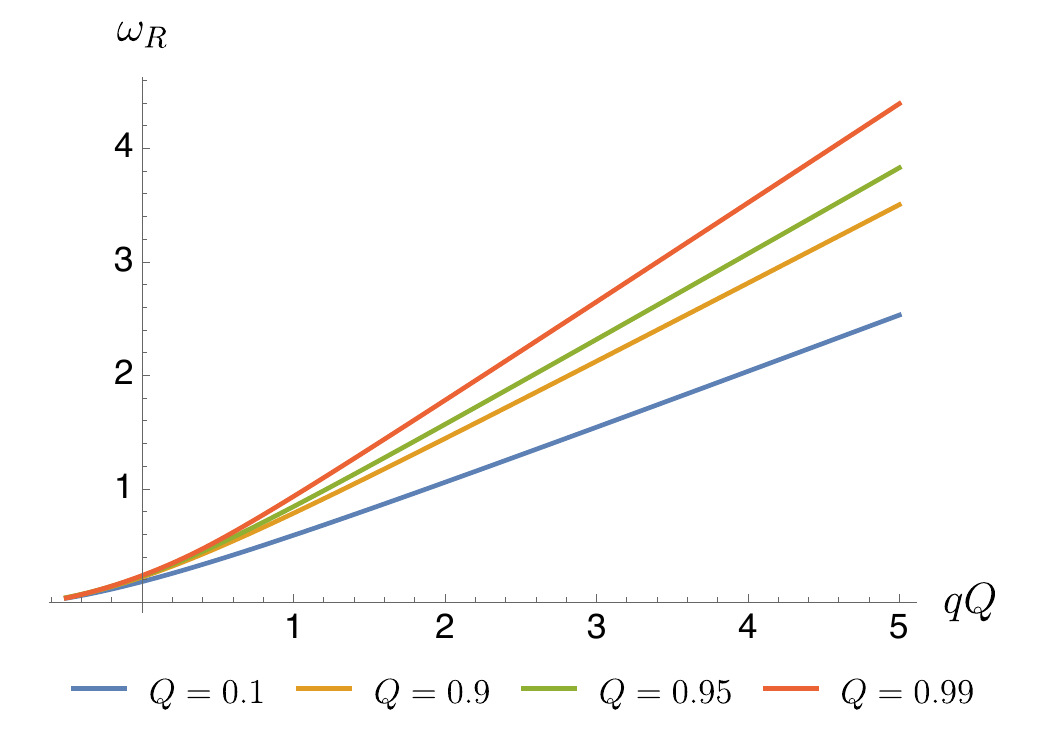}& & \hspace{-3mm}\includegraphics[scale=0.25]{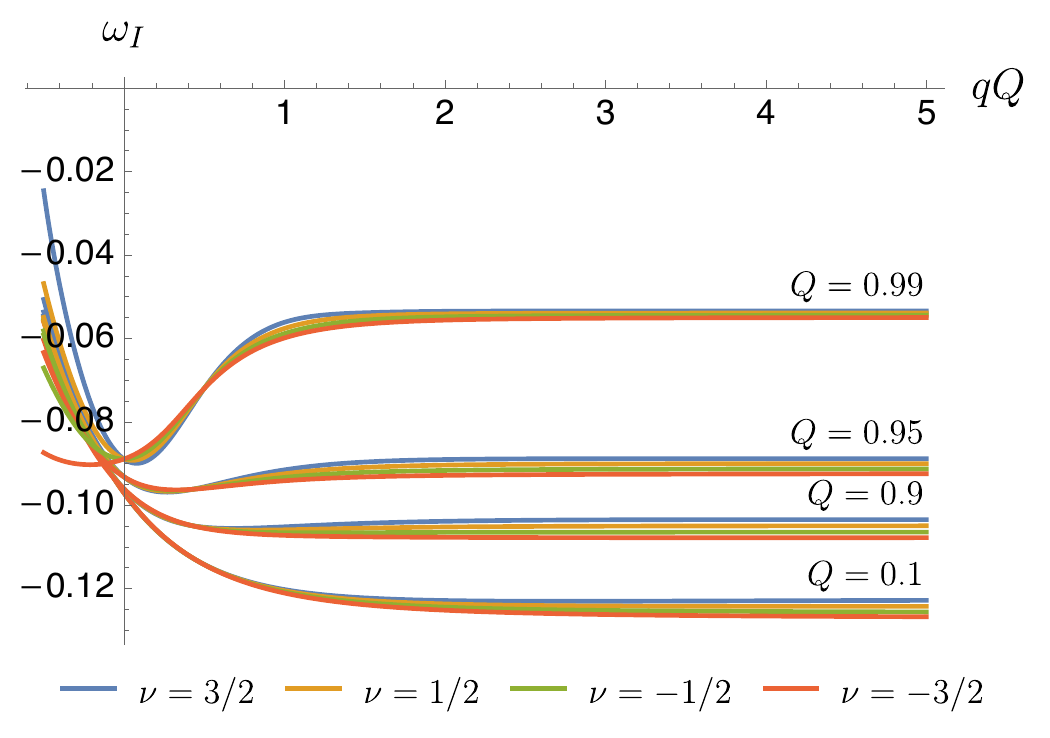}
\end{tabular} 
\end{center}


The final set of figures  illustrates  the profile of the NC fundamental QNM, by showing the $\omega_R- \omega_I$  plot parametrized by $Q/M$. The profiles are provided  for different projections $\nu$ with $Q/M$ ranging from $0.01$ to $0.96$. Left panel corresponds to the channel   $(j=1/2, s=1/2)$
and the right panel  corresponds to the channel $(j=3/2, s=1/2)$. It is clearly seen that the impact of noncommutativity grows with increasing $Q/M,$ as the lines corresponding to different projections   $\nu$ become more and more separated.

\begin{center}
\begin{tabular}{lll}
\includegraphics[scale=0.25]{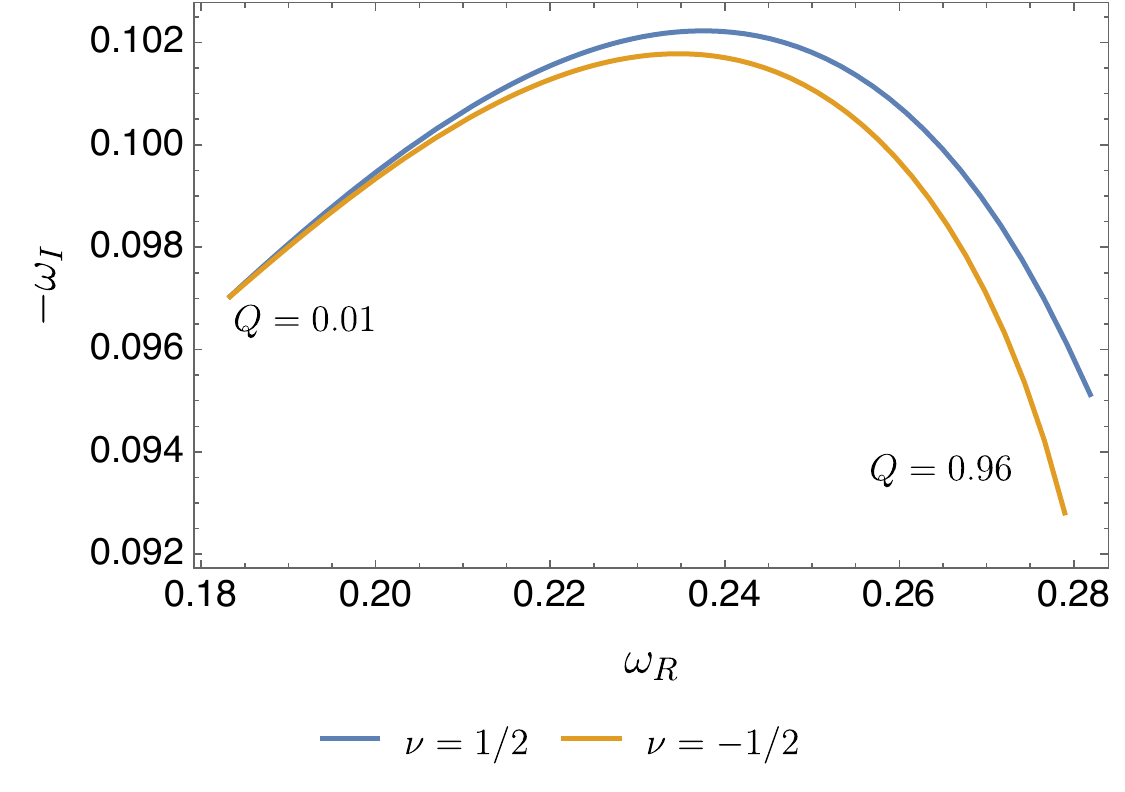}& & \hspace{-3mm}\includegraphics[scale=0.25]{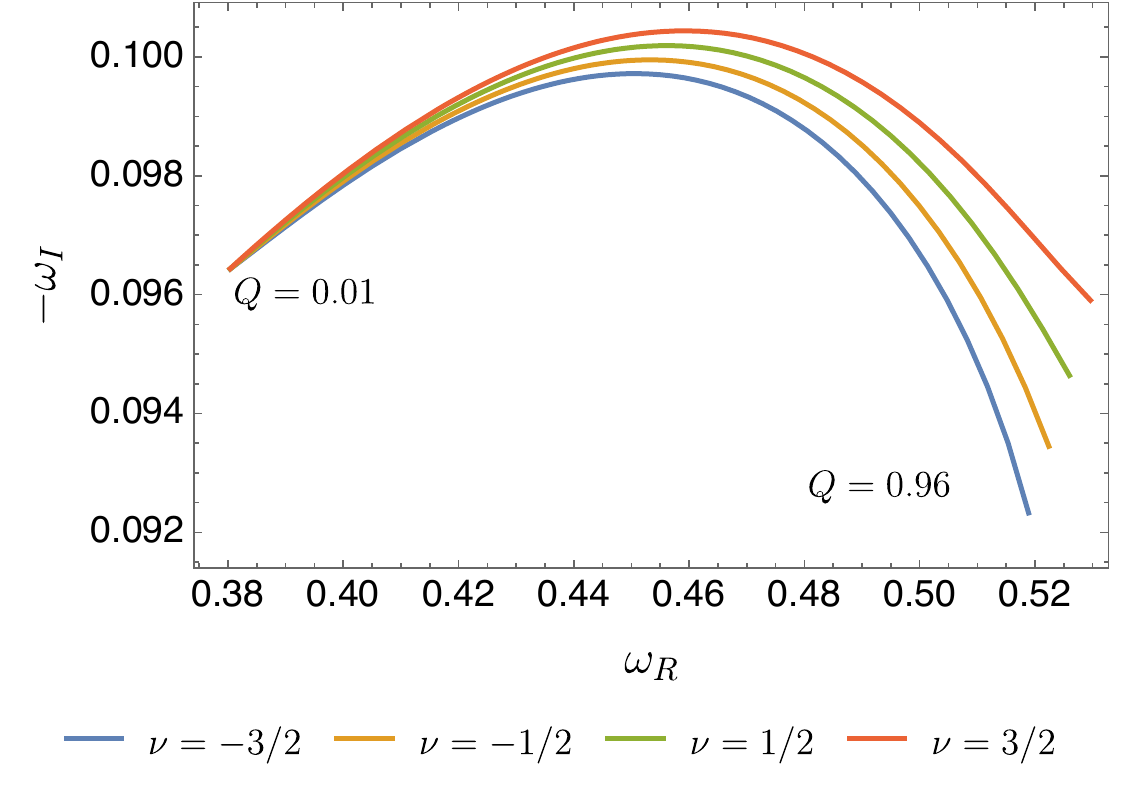}
\end{tabular} 
\end{center}



In this report we have not considered the massive Dirac perturbations. We plan to 
 address   them in the upcoming work   and analyze the corresponding  QNM spectrum. 
We also plan to study gray body factors for fermion particles that depend on the black's hole geometry, the fermion's mass, spin
and energy, as well as the specific gravity theory used to describe black holes.

\ack
This  research was supported by the Croatian Science
Foundation Project No. IP-2020-02-9614 \textit{Search for Quantum spacetime in Black Hole QNM spectrum and Gamma Ray Bursts}.
The work of  M.D.C. and  N.K.  is supported by Project 451-03-136/2025-03/200162 of the Serbian Ministry of Science, Technological
Development and Innovation.

\end{document}